\documentclass[cits]{PoS}
\usepackage{wrapfig}
\usepackage{multirow}
\usepackage{Tabbing}

\usepackage{caption}

\def\figurename{\footnotesize{Fig.}}

\def \eg {\it{e.g.}\rm~}

\title{ 
The COSPIX mission: focusing on the energetic and obscured Universe
}

\ShortTitle{COSPIX}

 \author{
 P.\,Ferrando$^{1,2}$, A.\,Goldwurm$^{1,2}$, P.\,Laurent$^{1, 2}$, O.\,Limousin$^{1,3}$, V.\,Beckmann$^{2,4}$, 
      M.\,Arnaud$^{1,3}$, X.\,Barcons$^5$, D.\,Bomans$^6$, I.\,Caballero$^{1, 3}$, F.~Carrera$^5$, S.\,Chaty$^{1, 3}$, J.\,Chenevez$^7$, R.\,Chipaux$^8$,
       F.~Christensen$^7$, A.\,Claret$^{1,3}$, S.~Corbel$^{1,3}$, J.~Croston$^9$, E.\,Daddi$^{1,3}$, M.\,De\,Becker$^{10}$, A.~Decourchelle$^{1,3}$, 
       D.\,Elbaz$^{1,3}$, M.~Falanga$^{11}$, C.~Ferrari$^{12}$, C.~Feruglio$^{1,3}$, D.\,G\"otz$^{1,3}$, C.\,Gouiff\`es$^{1,3}$, C.\,Hailey$^{13}$, 
       M.\,Hernanz$^{14}$, I.\,Kreykenbohm$^{15}$, J.\,Malzac$^{16}$, J.\,Martignac$^{1,3}$, F.\,Mattana$^2$, A.\,Meuris$^{1,3}$, G.\,Miniutti$^{17}$,
       K.\,Nalewajko$^{18}$, I.\,Negueruela$^{19}$, S.\,O'Dell$^{20}$, S.\,Paltani$^{21}$, R.\,Petre$^{22}$, P.-O.\,Petrucci$^{23}$, M.\,Pierre$^{1,3}$,
       F.\,Pinsard$^{1,3}$, G.\,Ponti$^9$, G.\,Rauw$^{10}$, N.\,Rea$^{14}$, M.\,Renaud$^{24}$, J.-L.\,Robert$^2$, J.\,Rodriguez$^{1,3}$, A.\,Rozanska$^{18}$
       A.\,Santangelo$^{25}$, J.-L.\,Sauvageot$^{1,3}$, S.\,Soldi$^{1,3}$, M.\,Tagger$^{26}$, C.\,Tenzer$^{25}$, R.\,Terrier$^2$, G.\,Trap$^{1,2}$, 
       P.\,Varni\`ere$^2$, J.\,Wilms$^{15}$, W.\,Zhang$^{22}$, J.\,Zurita Heras$^{2,4}$
       \\
       $^1$DSM/Irfu/SAp CEA Saclay FR; 
       $^2$APC Paris FR; 
       $^3$AIM Saclay FR; 
       $^4$FACe Paris FR;
       $^5$IFCA-CSIC-UC Santander ES; 
       $^6$Astr.\,Inst.\,Bochum DE;
       $^7$DTU Space Copenhagen DK; 
       $^8$DSM/Irfu/SEDI CEA Saclay FR; 
       $^9$Univ.\,Southhampton UK; 
       $^{10}$Inst.\,Astr.\,Geo\,Li\`ege BE;\\
       $^{11}$ISSI Bern CH;
       $^{12}$OCA Nice FR;
       $^{13}$Col.\,Univ.\,New-York USA;
       $^{14}$ICE/CSIC Barcelona ES;
       $^{15}$Ast.\,Inst.\,Univ.\,Erl.\,Bamberg DE;
       $^{16}$IRAP Toulouse FR;
       $^{17}$CAB/CSIC-INTA Madrid ES;
       $^{18}$Nic.\,Cop.\,Astr. Ctr.\,Varsaw PL;
       $^{19}$Dept.\,Fisica Univ.\,Alicante, ES;\\
       $^{20}$NASA/MSFC Huntsville USA;
       $^{21}$ISDC Gen\`eve CH;
       $^{22}$NASA/GSFC Greenbelt USA;\\
       $^{23}$LAOG Grenoble FR;
       $^{24}$LPTA Montpellier FR;
       $^{25}$IAAT T\"ubingen DE;
       $^{26}$LPC2E Orl\'eans FR
       \\
         E-mail: \email{philippe.ferrando@cea.fr}}

\abstract{Tracing the formation and evolution of all supermassive black holes, including the obscured ones, understanding how black holes influence their surroundings and how matter behaves under extreme conditions, are recognized as key science objectives to be addressed by the next generation of instruments. These are the main goals of the COSPIX proposal, made to ESA in December 2010 in the context of its call for selection of the M3 mission. In addition, COSPIX, will also provide key measurements on the non thermal Universe, particularly in relation to the question of the acceleration of particles, as well as on many other fundamental questions as for example the energetic particle content of clusters of galaxies.

COSPIX is proposed as an observatory operating from 0.3 to more than 100\,keV. The payload features a single long focal length focusing telescope offering an effective area close to ten times larger than any scheduled focusing mission at 30\,keV, an angular resolution better than 20 arcseconds in hard X-rays, and polarimetric capabilities within the same focal plane instrumentation. 

In this paper, we describe the science objectives of the mission, its baseline design, and its performances, as proposed to ESA.}

\FullConference{25$^{th}$ Texas Symposium on Relativistic Astrophysics - TEXAS 2010\\
		December 06-10, 2010\\
		Heidelberg, Germany}

\begin{document}
\section{Introduction}
\vspace{-0.5\baselineskip}
The opening of the X and gamma-ray windows in the sixties, thanks to satellite borne instruments, has started a new era in astrophysics. Extremely energetic and violent phenomena were discovered and found to be ubiquitous in the Universe, and the existence of compact objects, in particular of black holes, was unveiled. Since then the importance of the role of the black holes of all masses in the Universe has been realized to a point that their study is central in two themes of the Cosmic Vision science objectives of ESA, ``matter under extreme conditions'', and ``the evolving violent Universe''.

Observations in the hard X--ray range are crucial for these research themes. This domain is presently covered by simple imagers (as IBIS/ISGRI on the ESA INTEGRAL mission), orders of magnitude less sensitive and accurate than the powerful focusing instruments in soft X-rays (as XMM--Newton). The situation should change dramatically with the launch of the American NuSTAR \cite{Harrison10} and the Japanese Astro--H \cite{Taka10} missions, with focusing telescopes in hard X--rays. But their angular resolution (45'' and 100'' resp.) and effective area ($\sim$ 300\,cm$^2$ at 30\,keV) are still modest. If these missions will undoubtedly allow substantial progress in the field, it is already known that they will leave some questions (as \eg the Cosmic X-ray Background origin) only partly answered, and some others (as \eg the understanding of SgrA*) untouched in practice.

The Compact Object Spectroscopy Polarimetry and Imaging in hard X-rays (COSPIX) mission is designed to overcome the limitations of these future missions, in particular in terms of sensitivity and confusion limit. COSPIX is proposed as an observatory type mission, operating from below 0.5\,keV to about 100\,keV, with a very large effective area ($\sim$\,2000\,cm$^2$ at 30\,keV) an excellent angular resolution (<~20"), and polarimetric capabilities. Such an instrument will not only guarantee that fundamental scientific questions open today are fully addressed, but it will also allow in depth studies of the new topics that are bound to appear with NuSTAR and Astro--H.

In the following, after giving the main scientific objectives of the mission and the corresponding requirements, we propose a possible mission concept, as well as a baseline payload definition, which extensively build on the developments and studies made these last years in the context of Simbol--X \cite{Ferrando09} and IXO.
  
\section{Scientific objectives and requirements}
\vspace{-0.5\baselineskip}
Understanding the accretion of matter onto compact objects, and particularly black holes, is of fundamental importance in astrophysics, as this phenomenon has profound implications on physics and cosmology. On physics, the black hole environment is the only place where we can observe General Relativity at play beyond the weak field limit. On cosmology, the history of accretion, imprinted in the Cosmic X-ray Background (CXB) is thought to have fundamental links with that of galaxy formation and evolution. We describe in slightly more details below the questions which are the main drivers of the COSPIX design. With COSPIX, we want to:\\
\noindent {\bf--} resolve in discrete sources more than 70\,\% of the CXB at its peak at $\sim$\,30\,keV and characterize these sources believed to be obscured Super Massive Black Holes (SMBH), not detected in soft X-rays. If the handful of objects in the current hard X-ray surveys, sensitive only to bright and local objects, account for less than  5\,\% of the CXB  \cite{Beckmann09}, these highly obscured sources might be common at high redshift ({\eg \cite{Gilli07}). To find them and to measure their obscuration is crucial for understanding the feedback of the SMBH activity on the galaxy evolution. Fig.\,\ref{CXB} shows a simulated COSPIX spectrum of a local Compton thick AGN, and of a QSO placed at z~=~1. With 100~ks COSPIX observations, it will be possible to determine the absorbing column density within 25~\% for AGNs moderately bright (L$_X = 10^{44}$~erg/s) at redshift 1, and to detect sources up to $z \sim 3$ down to L(2\,--\,10~keV) less than a few $10^{44}$~erg/s.\hfill
\begin{figure}[t]
 \includegraphics[clip=true, width=0.5\textwidth]{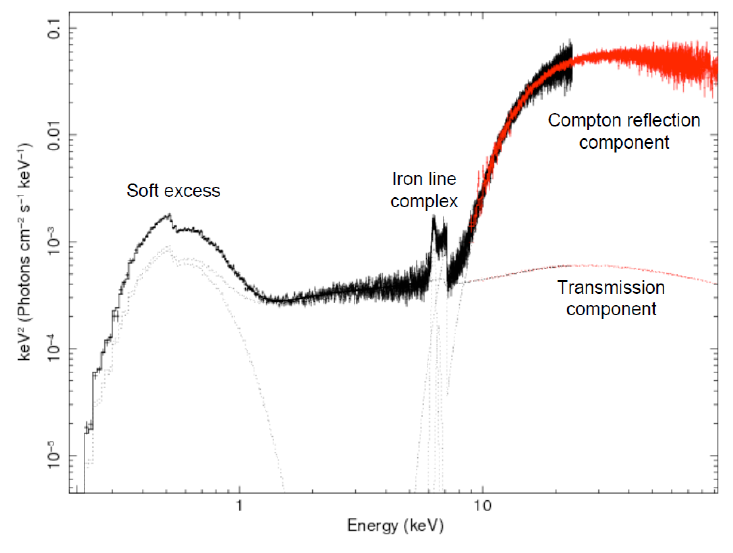}
 \includegraphics[clip=true, width=0.5\textwidth]{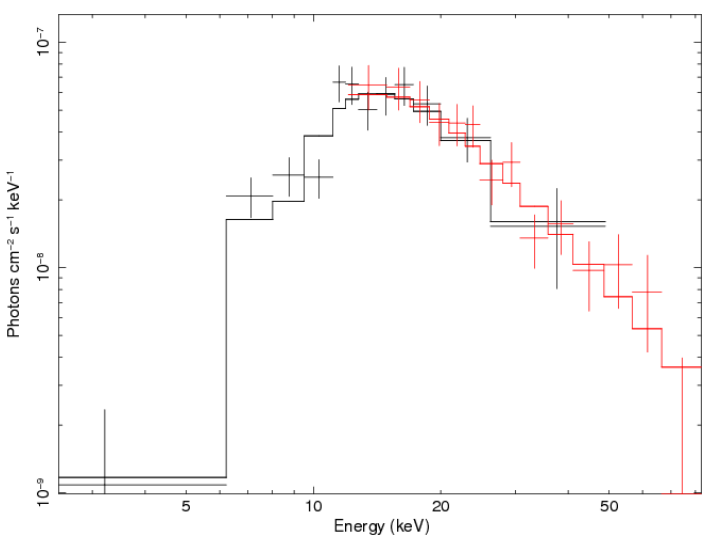}
  \vspace{-22pt}
 \caption{ {\it \bf left:} simulated spectrum of a 100~ks COSPIX observation of a local (z~=~0.02) Compton thick AGN with L(20\,--\,40~keV)~=~5\,10$^{43}$~erg/s; 
 {\it \bf right:} same for a 200~ks exposure of a Compton thick AGN like IRAS 09104+4109, placed at z~=~1, F(20\,--\,40~keV)~=~2\,10$^{-14}$~erg/s/cm$^2$. }
\label{CXB}
\vspace{-1.5\baselineskip}
\end{figure} \\
\noindent {\bf--} constrain accretion physics in all types of AGNs and galactic black holes, through the measurement of spectra in a broad energy range and with high statistics. COSPIX will make possible to perform time-resolved spectroscopy on the whole energy band on the dynamical time scale for a SMBH (a few ks for a 10$^7$ M$_\odot$ black hole), disentangling the different components (see Fig.\,\ref{CXB} for Seyferts) and looking for reverberation delays (\eg \cite{Fabian09}). With COSPIX this is more than 60 type I AGNs for which the reflection intensity will be measured at the \% level in 4\,ks integration times.
In addition to this classical spectral analysis, COSPIX will provide entirely new polarisation informations, in the hard X-ray part, allowing to assess the presence and role of synchrotron emission, challenging in particular the current paradigm of the blazar sequence.\hfill
\begin{wrapfigure}{t}{0.5\textwidth}
\vspace{-0.9\baselineskip}
\includegraphics[clip=true, trim=0.0cm 0.0cm 0cm 0cm, width=0.52\textwidth]{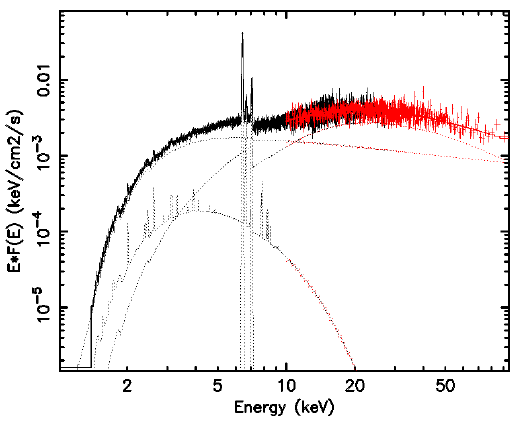}
\label{SgrA}
\vspace{-22pt}
\caption{simulation of a 100~ks COSPIX spectrum of G~011-011, a molecular cloud at 15' of SgrA*, with flux from \cite{Ponti10}.}
\vspace{-20pt}
\end{wrapfigure}
\noindent {\bf--} study the current and past activity of SgrA*. With its 4\,10$^6$\,M$_\odot$, this extremely dim black hole at the center of our galaxy provides a unique opportunity to study in details accretion on a SMBH, in order to understand distant AGNs, and the links between super massive and stellar mass black holes. The angular resolution and sensitivity of COSPIX will make it possible to study the puzzling flares of SgrA* (\eg \cite{Trap11}), despite the presence of strong close-by sources and the bright diffuse environment. In addition, if SgrA* is very faint today, it likely underwent a giant outburtst a few hundred years ago, with a luminosity 10$^6$ larger than present, as imprinted in the X-ray emission of close by Molecular Clouds \cite{Ponti10}. With COSPIX it will be possible to follow the variability of this reflected continuum on clouds (Fig.\,\ref{SgrA}), and through polarization measurements to definitively prove this hypothesis \cite{Churazov02} and provide a 3D localisation of these clouds. This will be the key to trace the past activity of SgrA* at times when it behaved as a low luminosity AGN.\\
\noindent {\bf--} measure the black hole spin distribution in a large population of objects. In the case of SMBH, the spin encodes the accretion history, and measuring its distribution will shed an entirely new light on the black hole growth and evolution. For galactic black holes, this carry information on the progenitor collapse and the black hole formation. The Fe line profile, from which the spin is determined, will be very well measured thanks to the COSPIX broad band removing uncertainties on the underlying continuum, and its good spectral resolution. With the COSPIX sensitivity, the black hole spin will be constrained to better than 30\,\% for about 100 Seyferts.
\begin{table}[t]
\centering
\caption[]{COSPIX main top level scientific requirements}
\vspace{-0.2\baselineskip}
\begin{tabular}{lll} 
\hline
Parameter & \multicolumn{2}{l}{Requirement} \\ 
\hline \hline
Energy range                                     & \multicolumn{2}{l}{ $\le$\,0.3 to >\,80\,keV }\\
On axis continuum sensitivity         
                                                 & $\le$\,2\,10$^{-15}$ c.g.s.  & 10\,--\,40\,keV, 3\,$\sigma$ in 1\,Ms \\
On axis line sensitivity @\,68\,keV 
                                                 & <\,10$^{-7}$\,ph/cm$^2$/s  &  3\,$\sigma$ in 1\,Ms \\
Minimum Detectable Polarisation  & \multicolumn{2}{l}{ <\,0.7\,\% \hspace{2em} 20\,--\,40\,keV, 3\,$\sigma$ in 100\,ks, 100\,mCrab source}\\
\hline
\multirow{2}{4cm}{On axis effective area}  
                                                  &  $\ge$\,2000\,cm$^2$ @ 1\,keV    &  $\ge$\,3500\,cm$^2$ @ 10\,keV\\
                                                  &  $\ge$\,1800\,cm$^2$ @ 30\,keV  &  $\ge$\,\hspace{0.5em}600\,cm$^2$ @ 75\,keV\\
\hline
Angular resolution (H.E.W)   & $\le$\,10'' @ E\,<\,10\,keV & $\le$\,20'' (goal 15) @ 30\,keV\\
Field of view @ 30\,keV         & diameter $\ge$ 6 arcmin (goal 9)\\
Absolute pointing accuracy  &  \multicolumn{2}{l}{ 3'' (goal 2) in radius, 90\,\%, after on-ground reconstruction} \\
\hline
Spectral resolution                 & $\Delta$E\,$\le$\,150\,eV @ 6 keV & $\Delta$E\,$\le$\,1 keV @ 68 keV \\
\hline
Time resolution                       & \multicolumn{2}{l}{ 50\,$\mu$s }\\
Asolute timing accuracy        & \multicolumn{2}{l}{ 100\,$\mu$s (goal 50) }\\
\hline 
Mission duration                     & \multicolumn{2}{l}{ 3 years (goal 5) after commissioning and calibration}\\
Total number of pointings    & >\,2000 (for 3 years) \\
\hline
\label{Req}
\end{tabular}
\vspace{-1.2\baselineskip}
\end{table}

The scientific requirements needed to reach these main science objectives are given in Table\,\ref{Req}. With these characteristics, COSPIX will also provide exquisite data on other important astrophysics studies, such as those of AGN jets and lobes (energetics, feed-back process), of neutron stars (equation of state, magnetic field), of clusters of galaxies (non thermal component), or of supernova explosion mechanism and nucleosynthesis ($^{44}$Ti line). It will also make a fundamental contribution to the study of particle acceleration in shocks, both relativistics (Pulsar Wind Nebulae), and non relativistics in a variety of physical parameters in colliding wind binaries and supernovae remnants (SNRs). In SNRs, COSPIX will provide a detailed mapping and spectral characterization of the accelerated electron population up to the cut-off, giving access to their maximum energy and putting strong constraints for the interpretation of the TeV emission as proton acceleration. 

\section{Scientific payload}
\vspace{-0.5\baselineskip}
In order to reach a high sensitivity in hard X-rays, there is no other way than implementing focusing optics, as presently done in soft X-rays, below $\sim$ 10\,keV. The instruments flying so far are using mirrors operated in total reflection, {\it i.e.\,} in small grazing incidence angles conditions. The low value of the largest angle for reflection to occur limits the energy range of such optics in the soft X-ray band for ``short'' focal lengths. An increase of the effective area in hard X-rays can be obtained either by increasing the focal length in total reflection regime or by using multilayers coatings which exploit Bragg diffraction and have the additional advantage to increase the field of view \cite{Pareschi08}. Using both, and a single optics, was the basis of the Simbol--X mission \cite{Ferrando09} which had a mirror area close to 400\,cm$^2$ at 30\,keV for a 20\,m focal length. An alternative path is taken by NuSTAR and Astro-H, which obtain similar, although slightly less,  areas with $\sim$~10\,m focal length optics only, by multiplying the number of telescopes. In practice however, this second approach cannot be used anymore to reach much larger areas, as those required in COSPIX, for which the only solution is to implement long focal length optics. The COSPIX scientific payload thus consists of a single telescope, made of an optics module focusing X-rays onto a focal plane detectors system.

\subsection{Optics payload}
\vspace{-0.25\baselineskip}
The main challenge of COSPIX is to build a large area and good angular resolution optics, with a limited mass, in order to fit within the M mission budget (and thus mass) constraints. The first point implies to use a Wolter I, or approximation to Wolter I, geometry, with a diameter over 1\,m, and a focal length of several tens of meters. The second point necessitates to use low mass density materials for the multilayers carriers, as done in the technologies of Silicon Pore Optics (SPO, \cite{Bavdaz10}) and Slumped Glass Optics (SGO, \cite{Zhang10}) developped in the context of the IXO mission. The COSPIX mirror takes advantage of both.
\begin{figure}[b]
\vspace{-1.0\baselineskip}
\begin{minipage}[t]{0.48\linewidth}
\includegraphics[clip=true, width=0.9\hsize]{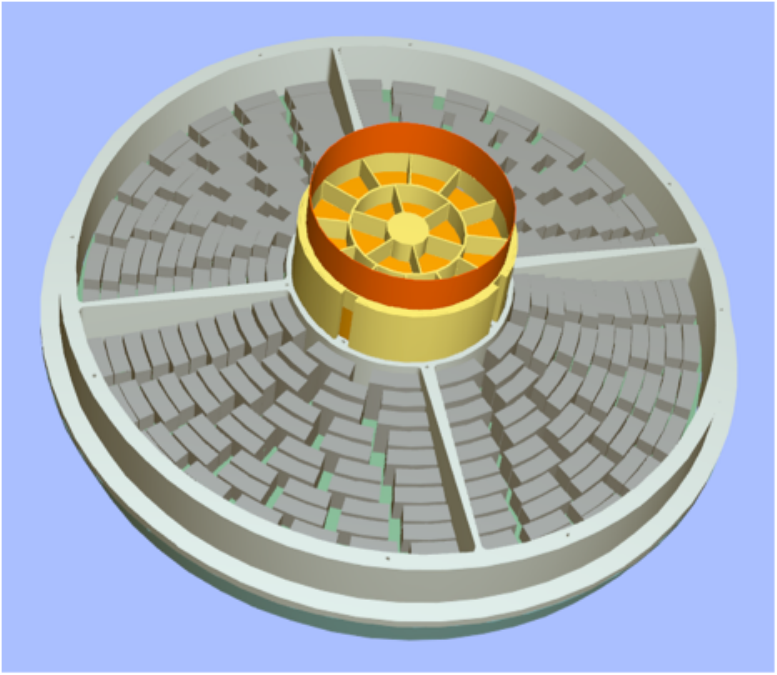}
    \caption{concept of COSPIX optics assembly,\\ combining SPO and SGO elements. \label{Optics}}
 \end{minipage} 
 \begin{minipage}[t]{0.5\linewidth}
    \hspace{-0.6cm} \includegraphics[clip=true, trim=0.2cm 0.4cm 0cm 0cm, width=1.1\hsize]{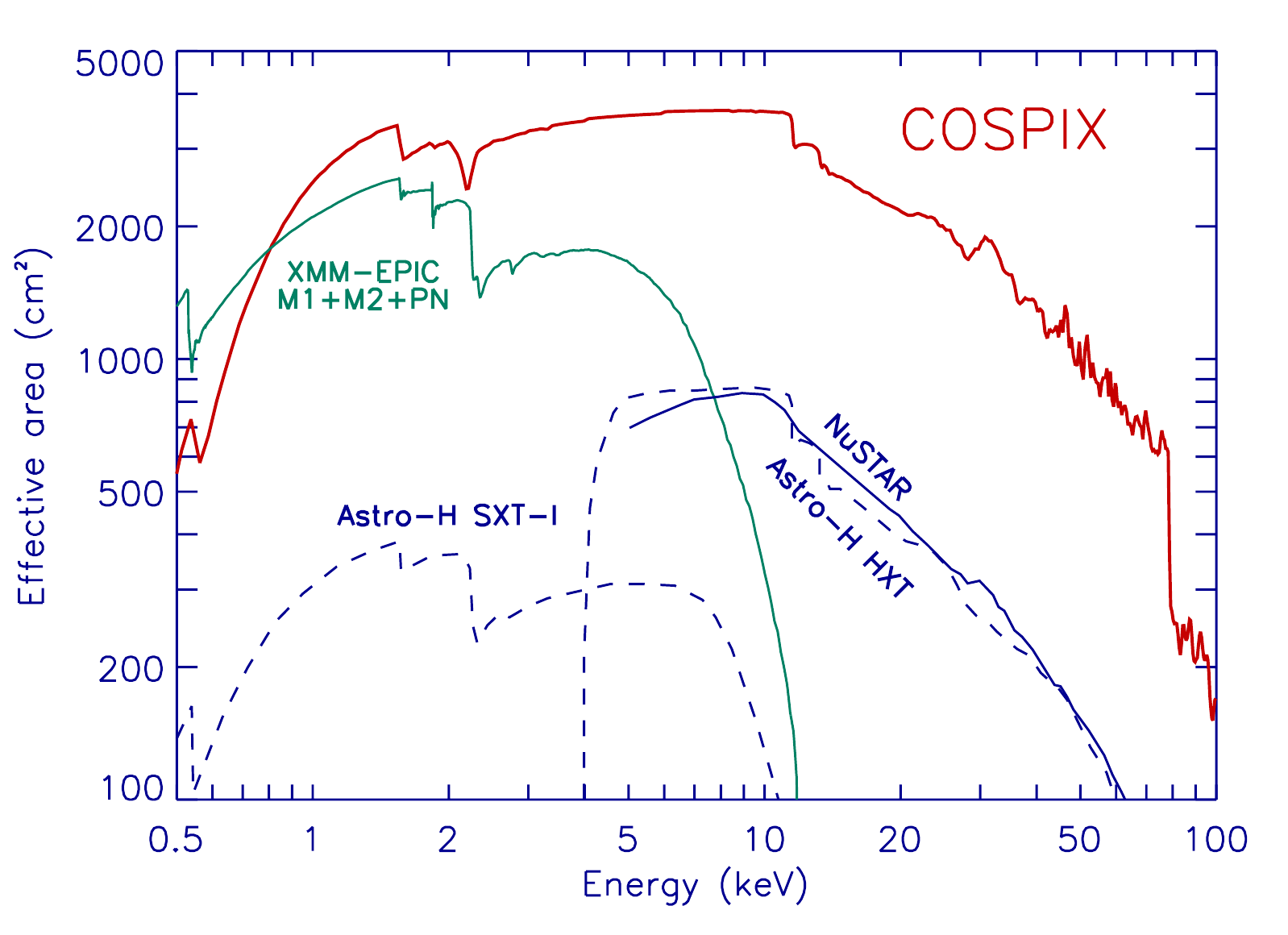}
     \caption{effective areas of COSPIX, Astro--H HXT and SXT-I, NuSTAR, and XMM--EPIC (all 3 cameras summed). \label{Area}}
     \end{minipage}
\end{figure}

Fig.\,\ref{Optics} shows the optics module design, with a focal length of 33\,m allowing to meet the area requirements. The SPO presents the best mass to area ratio with an excellent angular resolution and constitutes the outer part of the module. It spans from 50\,cm in diameter, a safe limit for silicon bending,  to 112\,cm, providing a large area for energies up to $\sim$~30\,keV. The highest energies are taken care by an SGO module inserted in the SPO, down to a diameter of 10\,cm. Both mirrors are equipped with sieve plates to minimize the straylight. This combination offers an area over 3500\,cm$^2$ at 10\,keV, and close to 650\,cm$^2$ at 75\,keV (Fig.\,\ref{Area}, with detection efficiencies included). The technological development and mounting processes are similar to those for IXO (\cite{Bavdaz10}, \cite{Zhang10}), ensuring that the COSPIX requirements, less stringent than for IXO, are met. The overall mass of the optics is 394\,kg, including the support structure and margins.

\subsection{Detector payload}
\vspace{-0.25\baselineskip}
The COSPIX focal plane design is completely inherited from the Simbol--X detector payload phase A studies \cite{Laurent09}, with a moderate scaling for the different geometrical conditions. It is made of two superimposed spectro-imaging devices, the Low Energy Spectro Imager (LESI) and the High Energy Spectro Imager (HESI), embedded in an an active / passive shielding system as shown in Fig.\,\ref{Detec}. This is complemented by a calibration wheel. The energy at which 50\,\% of the photons are stopped in the LESI is 17\,keV.

The LESI, which operates from 0.3 to 40\,keV, consists in a single $10\times10\rm~cm^2$ monolithic imager with a double-sided process. At the backside is a DEPFET active pixel sensor array, composed of 32768 pixels with a 520\,$\mu$m pitch, with a design fully similar to that of Simbol--X \cite{Lechner09}. At the front side the entrance window is divided in 32 square shaped pixels, with a 16.6\,mm pitch, with an effective threshold of $\sim$ 10\,keV; this is used for polarisation measurements on photons having a Compton interaction in the LESI and stopping in the HESI. The HESI, which operates from 8 to 250\,keV, has a design fully similar to that of Simbol--X \cite{Meuris09}. It is made of and array of $8\times8$ Caliste modules, and also covers a $10\times10\rm~cm^2$ area. Each Caliste module is equipped with a 2\,mm thick CdTe crystal having $16\times16$ pixels of 780\,$\mu$m pitch. The pixel size for the LESI ensures a 3-fold sampling of the optics point spread function at 10\,keV (4-fold for the HESI at 30\,keV). The LESI and HESI are cooled at moderately low temperatures ($-50^{\circ}$C and $-40^{\circ}$C resp.), sufficient to guarantee the required spectral resolution all along the mission lifetime. 

\begin{wrapfigure}{t}{0.5\textwidth}
\vspace{-0.4\baselineskip}
\includegraphics[width=\hsize, clip=true, trim=0.65cm 0cm 0cm 0cm]{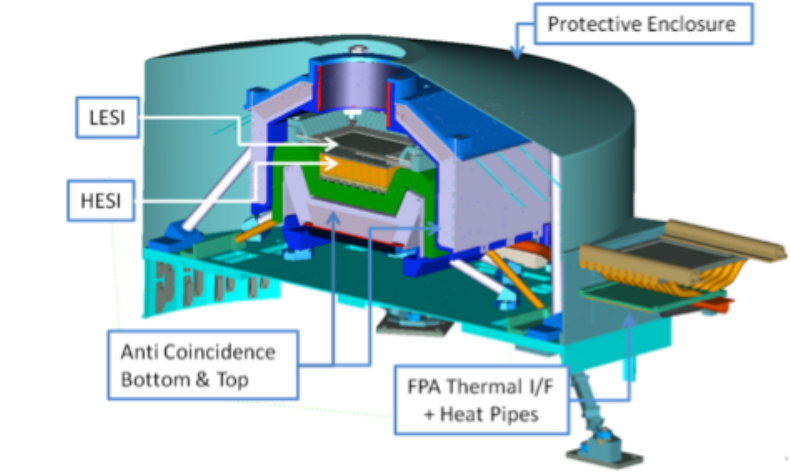}
\vspace{-18pt}
\caption{the COSPIX focal plane assembly. \label{Detec}}
\end{wrapfigure}
The low background required is achieved by combining two systems identical to those of Simbol--X \cite{Laurent09}, apart from a simple scaling. At the focal plane level the active shielding \cite{Chabaud09}, used in anticoincidence with the imaging detectors, reduces the background induced by particles. 
At the system level, the combination of a 3.1\,m long collimator on top of the detector assembly, made with graded shield material \cite{Laurent09}, and of a 3.5\,m and 33\,kg circular sky shield around the optics payload, ensures that no sky out of the mirror field of view is seen by the focal plane. The overall detector payload has a mass of 176\,kg (with collimator and structure), and a power consumption of 167\,W, margins included.
\begin{figure}[t]
\includegraphics[clip=true, trim=0.4cm 0.4cm 0cm 0cm, width=0.5\hsize]{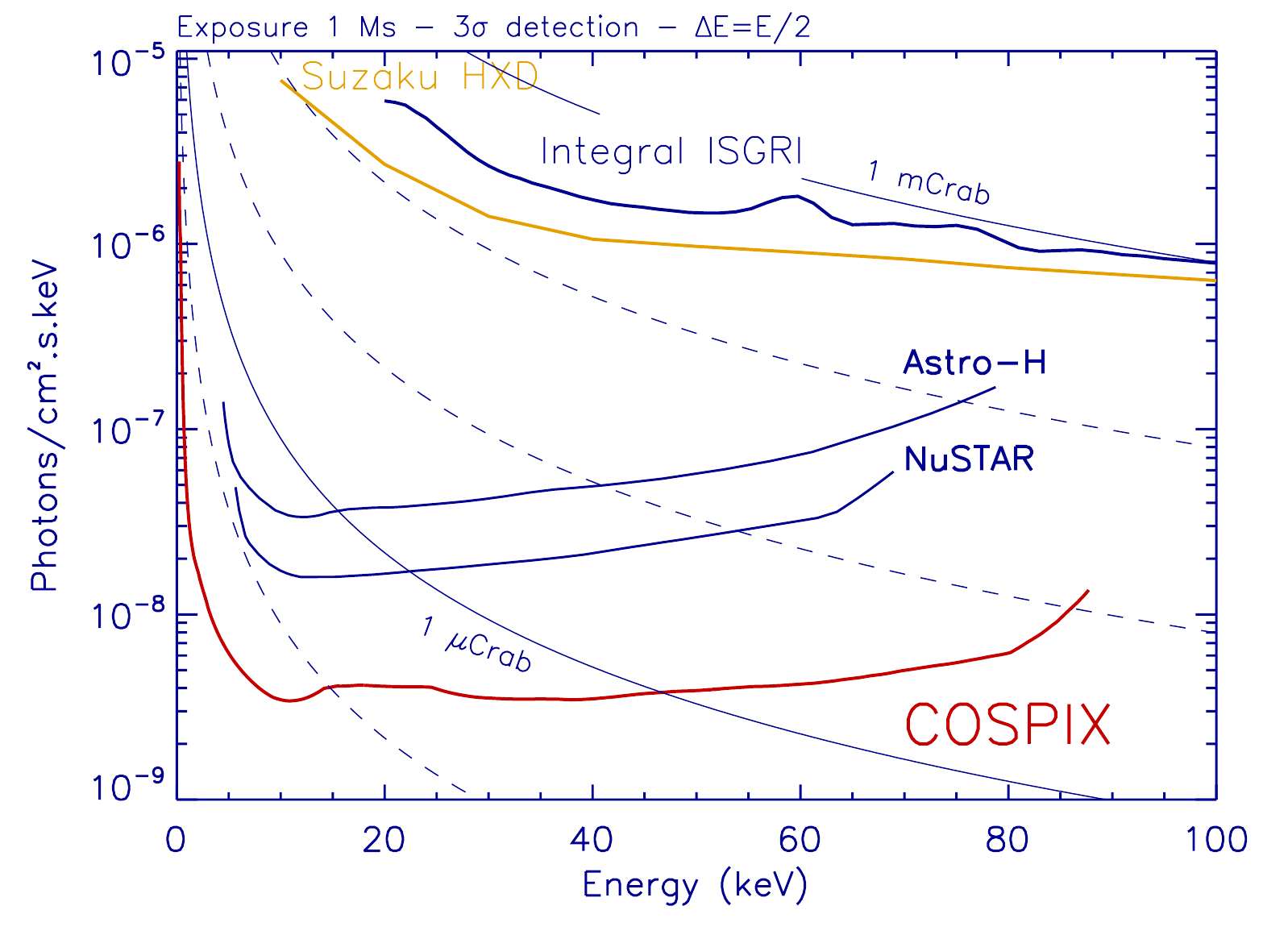}
\includegraphics[clip=true, trim=0.4cm 0.4cm 0cm 0cm, width=0.5\hsize]{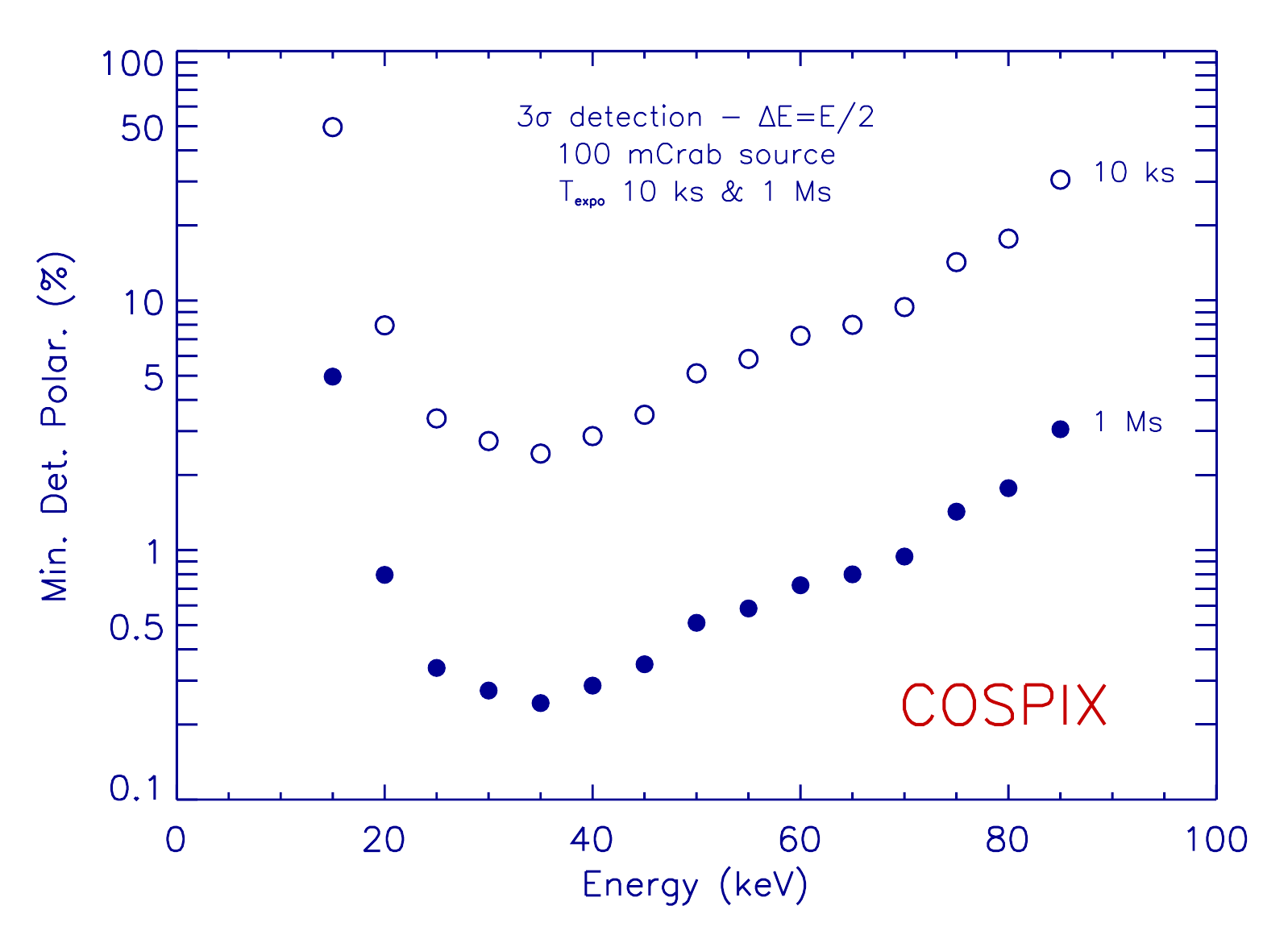}
\vspace{-20pt}
\caption{
 {\it \bf left:} continuum sensitivities of COSPIX for point sources, compared to flying missions and to the forthcoming NuSTAR and Astro--H missions (all evaluated for extraction region of PSF size, from published instrumental data); \\ {\it \bf right:} MDP of COSPIX, calculated for a 100\,mCrab intensity source, and for integration times of 10\,ks and 1\,Ms.}
\label{Perf}
\vspace{-15pt}
\end{figure}

\vspace{-1.0\baselineskip}
\section{Performances and mission implementation}
\vspace{-0.5\baselineskip}
The performances of the above model payload are fully consistent with the mission requirements (Table\,\ref{Req}), as shown in particular in Fig.\,\ref{Area} for the effective area, and in Fig.\,\ref{Perf} for the point source detection sensitivity and for the minimum detectable polarisation.

Regarding the payload implementation, and particularly the 33\,m focal length, industrial studies have shown that in can be achieved with a simple evolution of the Simbol--X design, namely two spacecrafts in a formation formation flight configuration, with the mirror on one spacecraft and the detector on the second one, slaved in position with respect to the mirror axis \cite{Cledassou05}. The requirements on the formation control and attitude reconstruction are very similar to those of Simbol--X, and can be met with the same systems, for metrology and propulsion.

It is proposed to place COSPIX in orbit around L2,  advantageous from several points of view with respect to a High Elliptical Orbit, starting with formation flight and thermal stability. The total launch mass is 2.06 tons (including over 30\,\% of margins and contingency) with consumables for five years, a mass suitable for a launch by a Soyuz rocket from Kourou. COSPIX can perform observations autonomously, with data recorded on board before download to the ground. For this, it is anticipated to use a single 15\,m antenna  in X band, with a rate of 1.6\,Mbps (corresponding to 1\,Crab intensity without compression), and for typically four hours every day.

COSPIX will be operated as an observatory with all standard aspects of observations selection and data distribution. The COSPIX development relies on a solid technical basis,  all elements being on a development track to reach TRL~5 well before the end of 2014. COSPIX can be ready for a launch late 2020.

\vspace{-0.5\baselineskip}

\end{document}